\documentclass[aps,prl,reprint]{revtex4-1}
\usepackage{threeparttable}
\usepackage[dvips]{graphics}
\usepackage{longtable}
\usepackage{enumerate}
\usepackage{amsfonts}
\usepackage{amsmath}
\usepackage{stmaryrd}
\usepackage{dcolumn}
\usepackage{xcolor}
\usepackage{bm}
\usepackage{graphicx}

\begin{document}
\title{Calculations of Time-Reversal Symmetry Violation Sensitivity Parameters Based on Analytic Relativistic Coupled-Cluster Gradient Theory}

\author{Chaoqun Zhang}
\affiliation{Department of Chemistry, The Johns Hopkins University, Baltimore, MD 21218, USA}

\author{Xuechen Zheng}
\affiliation{Department of Chemistry, The Johns Hopkins University, Baltimore, MD 21218, USA}

\author{Lan Cheng}
\email{lcheng24@jhu.edu}
\affiliation{Department of Chemistry, The Johns Hopkins University, Baltimore, MD 21218, USA}

\begin{abstract}

   We develop an analytic-gradient based method for 
   relativistic coupled-cluster calculations of effective electric field, $\mathcal{E}_{\text{eff}}$, with improved efficiency and robustness over the previous state of the art. The enhanced capability to calculate this time-reversal symmetry violation sensitivity parameter enables efficient screening of candidate molecules for the electron electric dipole moment (eEDM) search.
   As examples, 
   the $|\mathcal{E}_{\text{eff}}|$ values of metal methoxides including BaOCH$_3$, YbOCH$_3$, and RaOCH$_3$ are shown to be as large as those of the corresponding fluorides and hydroxides, which supports the recent proposal of using 
   these symmetric-top molecules to improve the sensitivity of eEDM measurements. The computational results also show that molecules containing late actinide elements, NoF, NoOH, LrO, and LrOH$^+$, exhibit particularly large $|\mathcal{E}_{\text{eff}}|$ values of around 200 GV/cm.

\end{abstract}

\maketitle

 \section{Introduction} 
  
 The numerous discoveries at the Large Hadron Collider (LHC) of CERN included the observation of the Higgs boson, which completed the search for the fundamental particles in the Standard Model \cite{Aad12}. The exploitation of the 14 TeV collision energy of LHC has not observed fundamental particles associated with physics Beyond the Standard Model. 
 One increasingly powerful method to search for new physics beyond the Standard Model is tabletop low-energy experiments based on precision spectroscopy of atoms and molecules \cite{DeMille17,Safronova18,Cairncross19}. In particular, the search for electron electric dipole moment (eEDM) through precision measurements of paramagnetic atoms and molecules has emerged as a promising route \cite{DeMille17,Cairncross19}.
 
 In paramagnetic atoms and molecules, the interaction between the eEDM (d$_e$) and the effective electric field, d$_e\mathcal{E}_{\text{eff}}$, contributes to the atomic and molecular energy levels that are subject to spectroscopic interrogation. Although no nonzero eEDM has been reported, these measurements have set upper bounds to the eEDM value, which provides valuable information about the lower bounds for the energies of new fundamental particles. The sensitivity of the eEDM measurements is directly proportional to the electric field that the electrons experience. Paramagnetic atoms and molecules are sensitive to measurements of the eEDM because the $|\mathcal{E}_{\text{eff}}|$ values are far greater than applied laboratory electric fields \cite{Sandars65}. Furthermore, since paramagnetic molecules exhibit orders of magnitude larger $|\mathcal{E}_{\text{eff}}|$ values than atoms \cite{Sandars67}, the recent work on diatomic molecules including YbF \cite{Hudson02,Hudson11}, ThO \cite{Baron14,Andreev18}, and HfF$^+$ \cite{Cairncross17} has reduced the upper bound of eEDM by two orders of magnitude compared with a previous record set by the Thallium atom \cite{Regan02}. The present record of 1.1$\times10^{-29}$ $e \cdot \mathrm{cm}$ 
 for the upper bound of eEDM obtained from the measurements of ThO \cite{Andreev18} corresponds to an energy scale of around 30 TeV for certain classes of new fundamental particles, which is beyond the reach of LHC. The ongoing experiments 
 powered with new techniques to improve the precision \cite{Lim18,Panda19,Wu20,Zhou20,Verma20,Ho20} hold the promise to significantly improve the current limit. 
 Furthermore, new schemes to use nearly degenerate vibrational and rotational states in laser-cooled polyatomic molecules have the potential to enhance the sensitivity by another several orders of magnitude \cite{Kozyryev17,Augenbraun20,Hutzler20}. 
 
 The $|\mathcal{E}_{\text{eff}}|$ value, which represents the strength of an internal effective electric field in a paramagnetic atom or molecule, is not accessible to direct measurements.
 Electronic-structure calculations for this parameter \cite{kozlov1987calculation,dmitriev1992calculation,kozlov1997enhancement,parpia1998ab,quiney1998hyperfine,ravaine2005marked,Meyer06,Meyer08,Skripnikov13,Lee13,Abe14,Fleig14,Kudashov14,Skripnikov15,Skripnikov15a,Denis15,Prasannaa15,Sasmal15c,Fleig16,Skripnikov16,Sasmal16,Skripnikov17,Fleig17a,Abe18,Denis19,Prasannaa19,Gaul20,Gaul20a,Talukdar20,Mitra21} thus play an important role in the interpretation of experimental measurements and in the selection of candidate molecules. $\mathcal{E}_{\text{eff}}$ is a ``first-order property'', i.e., it corresponds to a first derivative of the electronic energy. Since $\mathcal{E}_{\text{eff}}$ samples the electron density in the core region, it requires accurate treatments of both relativistic and electron-correlation effects to obtain accurate 
 values. 
 The calculations of $\mathcal{E}_{\text{eff}}$ have relied on relativistic wave function based methods. 
 The available calculations have used relativistic coupled-cluster (CC) methods \cite{Abe14,Skripnikov13,Sasmal15c,Abe18} and multireference configuration interaction (CI) or CC methods \cite{Fleig14,Fleig17a}. 
 The calculations to date have used approximate unrelaxed formulations \cite{Skripnikov13,Abe14,Fleig14,Sasmal15c} or numerical differentiation of electronic energies \cite{Skripnikov16,Abe18,Denis19}. Although numerical differentiation can reproduce either relaxed or unrelaxed gradients, by including the finite perturbation in the Hartree-Fock (HF) or post-HF calculations, the numerical-differentiation procedure for obtaining $\mathcal{E}_{\text{eff}}$ is tedious due to the sensitivity of the numerical results to the step size and the convergence of energy calculations \cite{Abe18,Denis19}. These together with high computational cost of relativistic wave function calculations render the calculations of $\mathcal{E}_{\text{eff}}$ still a major challenge. 
 
 The tremendous efforts devoted to molecular structural optimization in quantum chemistry have established analytic gradients \cite{Pulay69,Pople79,Helgaker88,Stanton00,Pulay14} as the standard tool for the calculations of first-order molecular properties. For example, a single analytic CC gradient calculation, which is two to three times as costly as a corresponding energy calculation, provides all first-order properties \cite{Stanton00}. An analytic-gradient based scheme is not only by far more efficient than numerical differentiation of electronic energies, but is also convenient to use.
 In this Letter we report the development of an analytic-gradient based method for the calculations of $\mathcal{E}_{\text{eff}}$ using the relativistic exact two-component (X2C) CC 
 analytic-gradient theory, hereby combining the analytic X2C gradient theory \cite{Cheng14,Filatov14} and the recent development \cite{Liu21} of analytic first derivatives \cite{Scheiner87,gauss91a,Scuseria91,Watts93} for CC singles doubles (CCSD) \cite{Purvis82} and CCSD with a noniterative triples [CCSD(T)] \cite{Raghavachari89} methods with spin-orbit coupling included at the orbital level. The development of the present analytic-gradient based method aims to significantly improve the efficiency, robustness, and convenience for the calculations of $\mathcal{E}_{\text{eff}}$ to enable rapid and  reliable screening of candidate molecules for use in the eEDM measurements.

 \section{Theory}
 
  Relativistic electronic-structure calculations of effective electric field, $\mathcal{E}_{\text{eff}}$, are based on the Dirac Hamiltonian 
    \begin{eqnarray}
    \hat{H}=\hat{H}_0+ d_e \hat{V}_{\text{eff}},
    \end{eqnarray}
    with $\hat{H}_0$ and $d_e \hat{V}_{\text{eff}}$ representing the
    one-electron Dirac operator and the interaction between eEDM and
    the effective electric field \cite{Johnson86}
\begin{eqnarray}
    \hat{H}_0=
    \left( \begin{array}{cc}
        \hat{V} & c\vec{\sigma}\cdot \vec{p}  \\
        c\vec{\sigma}\cdot \vec{p} & \hat{V}-2c^2 
    \end{array} \right)~,~
	\hat{V}_{\text{eff}}
	=-2ic\beta\gamma_5 \hat{p}^2,
\end{eqnarray}
in which $c$ is the speed of light, $\vec{\sigma}$ is a vector of Paul spin matrices,
$\vec{p}$ is the momentum operator, $\hat{V}$ is
the nuclear attraction potential, and 
$\beta$ and $\gamma_5$ are Dirac matrices given by 
\begin{eqnarray}
    \beta = 
    \left( \begin{array}{cccc}
    1 & 0 & 0 & 0 \\
    0 & 1 & 0 & 0 \\
    0 & 0 & -1& 0 \\
    0 & 0 & 0 & -1\\
    \end{array} \right)~,~
    \gamma_5 = 
    \left( \begin{array}{cccc}
    0 & 0 & 1 & 0 \\
    0 & 0 & 0 & 1 \\
    1 & 0 & 0 & 0 \\
    0 & 1 & 0 & 0 \\
    \end{array} \right).
\end{eqnarray}
$\mathcal{E}_\mathrm{eff}$ corresponds to 
the first derivative of the electronic energy with respect to $d_e$
\begin{eqnarray}
    \mathcal{E}_\mathrm{eff} =
    \frac{\partial E}{\partial d_e}|_{d_e=0}.
\end{eqnarray}
Since $\hat{V}_{\text{eff}}$ involves the second derivatives
of the small component wave function, $\mathcal{E}_\mathrm{eff}$ samples the wave function 
in the core region and is sensitive to treatments of relativistic and electron-correlation effects. 
The wave functions of the electronic states 
used in eEDM measurements 
, e.g., the $X^2\Sigma$ state of YbF and the
$^3\Delta_1$ states of ThO, HfF$^+$, and ThF$^+$,
are dominated by a single electron configuration. 
CC methods \cite{Bartlett07, Crawford00}
can provide accurate  
treatments for dynamic correlation and 
are the methods of choice here. 

The present scheme for the calculations of $\mathcal{E}_\mathrm{eff}$ uses the recent implementation of analytic first derivatives for X2C CCSD and CCSD(T) methods \cite{Liu21} and also the atomic-orbital based algorithms 
\cite{Liu18b} to enhance the computational efficiency. 
The calculation of $\mathcal{E}_\mathrm{eff}$ using the X2C-CC analytic-gradient theory involves a simple contraction between the relaxed one-electron density matrix
$D^{\text{X2C-CC}}$ and the corresponding property integrals
$[V_\mathrm{eff}]^{\text{X2C}}$
\begin{eqnarray}
    \mathcal{E}_\mathrm{eff} =
    \frac{\partial E_{\mathrm{X2C-CC}}}{\partial d_e}|_{d_e=0}
	=\sum_{pq} D^{\text{X2C-CC}}_{pq} [V_\mathrm{eff}]^{\text{X2C}}_{pq}. 
\end{eqnarray}
We refer the readers to Ref. \cite{Liu21} for the calculations of the relaxed one-electron density matrix $D^{\text{X2C-CC}}$. 
We base the calculation of $[V_\mathrm{eff}]^{\text{X2C}}$ 
on the analytic X2C gradient theory
\cite{Cheng14,Filatov14}.
The X2C theory \cite{Dyall97, Kutzelnigg05, Liu09} uses the matrix representation of the one-electron Dirac equation 
\begin{eqnarray}
    \left( \begin{array}{cc}
	    h^{\text{LL}} & h^{\text{LS}}  \\
	    h^{\text{SL}} & h^{\text{SS}} 
    \end{array} \right)
    \left( \begin{array}{c}
	    C^{\text{L}}  \\
	    C^{\text{S}} 
    \end{array} \right)
    =
    \left( \begin{array}{cc}
	    S      & 0       \\
	    0      & \frac{T}{2c^2}
    \end{array} \right)
    \left( \begin{array}{c}
	    C^{L}  \\
	    C^{S} 
    \end{array} \right)
    E,
\end{eqnarray}
in which $C^{\text{L}}$ and $C^{\text{S}}$
are large- and small-component orbital coefficients
in kinetically balanced basis sets \cite{Stanton84}
\begin{eqnarray}
	\phi^{\text{L}}_i=C^{\text{L}}_{\mu i} f_\mu~,~
	\phi^{\text{S}}_i=C^{\text{S}}_{\mu i} \frac{\vec{\sigma}\cdot \vec{p}}{2c}f_\mu,
\end{eqnarray}
$h^{\text{LL}}$, $h^{\text{LS}}$, $h^{\text{SL}}$, $h^{\text{SS}}$
are the large-large, large-small, small-large, 
and small-small blocks of the Hamiltonian matrix
\begin{eqnarray}
h^{\text{LL}}_{\mu\nu}&=&V_{\mu\nu}~,~ 
h^{\text{LS}}_{\mu\nu}=h^{\text{SL}}_{\mu\nu}=T_{\mu\nu}, \\
h^{\text{SS}}_{\mu\nu}&=&\frac{1}{4c^2}\langle f_\mu|\vec{\sigma}\cdot\vec{p} V\vec{\sigma}\cdot\vec{p}|f_\nu\rangle-T_{\mu\nu}, 
\end{eqnarray}
and $S$, $T$, and $V$ represent overlap, kinetic energy, and nuclear attraction potential matrices. 
This four-component Hamiltonian matrix is block-diagonalized \cite{Dyall97}
\begin{eqnarray}
    \left( \begin{array}{cc}
	    h^{\text{LL}} & h^{\text{LS}}  \\
	    h^{\text{SL}} & h^{\text{SS}} 
    \end{array} \right)
\to  
\left( \begin{array}{cc}
	    h_+ & 0  \\
	    0 & h_-
    \end{array} \right)
\end{eqnarray}
to decouple electronic and positronic degrees of freedom.
The electronic block
\begin{eqnarray}
	&&h_+=R^\dag [h^{\text{LL}}+h^{\text{LS}}X+X^\dag h^{\text{SL}} + X^\dag h^{\text{SS}} X]R, \\
	&&C^{\text{S}}=XC^{\text{L}},R=(\tilde{S}^{-1}S)^{1/2},\tilde{S}=S+\frac{1}{2c^2}X^\dag TX,
\end{eqnarray}
is used together with the untransformed two-electron Coulomb interaction in the subsequent many-electron treatment. We obtain $[V_\mathrm{eff}]^{\text{X2C}}$  by differentiating
$h_+$, using a procedure developed in Ref. \cite{Cheng11b}. 
The calculation of $[V_\mathrm{eff}]^{\text{X2C}}$ 
involves the derivatives of $h^{\text{LS}}$ and $h^{\text{SL}}$,
since $V_\mathrm{eff}$ appears on the LS and SL blocks,
\begin{eqnarray}
\begin{scriptsize}
\frac{\partial h^{\text{LS}}_{\mu\nu}}{\partial d_e}
=\frac{\partial h^{\text{SL}\ast}_{\nu\mu}}{\partial d_e}
=\langle f_\mu|
    \left( \begin{array}{cc}
	    -2icp^2 & 0  \\
	    0 & -2icp^2 
    \end{array} \right)
|\frac{\vec{\sigma}\cdot \vec{p}}{2c}f_\nu\rangle,
    \end{scriptsize}
\end{eqnarray}
and the derivatives of the $X$ and $R$ matrices.

\section{Computational Details}
We have implemented the present analytic-gradient based method for the calculations of effective electric field in the CFOUR program package \cite{cfour,Matthews20CFOUR}
and used it for all the calculations presented here. 
Our calculations used experimental equilibrium structures for HfF$^+$ \cite{Cossel2012}, TaN \cite{Ram02}, ThF$^+$ \cite{Gresh16}, ThO \cite{Edvinsson84}, BaF\cite{Knight1971}, BaOH \cite{KinseyNielsen1986}, YbF \cite{Dickinson2001}, YbOH \cite{Nakhate2019}, a relativistic Fock-space CCSD bond length for HgF \cite{Knecht11} to enable direct comparison with Ref. \cite{Prasannaa15}, and 
the spin-free X2C \cite{Dyall01,Liu09, Cheng11b} CCSD(T)\cite{Raghavachari89}/cc-pVTZ \cite{Dunning89,Lu16,Feng17,Hill17} structures for the other molecules.
which are documented in detail in the Supporting Information \cite{SI}.

Unless otherwise stated, the calculations employed the X2C Hamiltonian \cite{Dyall97,Kutzelnigg05,Ilias07} with the atomic mean-field \cite{Hess96} spin-orbit integrals (the X2CAMF scheme)
\cite{Liu18} and Gaussian nuclear model \cite{Visscher97},
and included the Gaunt term
in the AMF approach.
We used large uncontracted basis sets and correlated valence
and semicore electrons. 
We mention that the basis-set errors beyond the uncontracted triple-zeta basis sets 
have been shown to be small \cite{Fleig14,Skripnikov15,Abe18,Denis19}. 
The present calculations employed the uncontracted ANO-RCC basis sets \cite{Faegri01,Roos05,Roos05a,Roos08} for heavy atoms, which are of augmented quadruple-zeta quality for
W, Ta, Ba, Ra, Hg, and Lu and of augmented triple-zeta quality
for Yb, Hf, and Th, except that we used the uncontracted cc-pVTZ basis sets \cite{Feng17} for 
No and Lr. For light elements, the uncontracted cc-pVTZ basis sets \cite{Dunning89} were used in calculations of
RaOH, YbOCH$_3$, BaOCH$_3$, RaOCH$_3$, LuOH$^+$, LrO, LrOH$^+$, NoF, NoOH, while the uncontracted aug-cc-pVTZ basis sets were used for the other molecules.
Although we used aug-cc-pVTZ basis sets in many calculations,
the contributions from diffuse functions to $|\mathcal{E}_\mathrm{eff}|$'s turned out to be negligible, e.g.,
they amount to less than 0.2\% for BaOH, YbOH, and RaOH.


\section{Results and Discussions}
  \begin{table}
  \begin{center}
  \caption{The $|\mathcal{E}_\mathrm{eff}|$ values (GV/cm) from the 
  X2CAMF-HF, CCSD, and CCSD(T) calculations using the analytic energy-gradient theory.
  ``fc ele'' refers to the number of core electrons kept frozen together with virtual orbitals
  higher than 100 Hartree in
  the CC calculations.}
  \label{tab1}
  \begin{tabular}{ccccccccc}
    \hline \hline
    & fc ele & HF & CCSD & CCSD(T) & Literature \\
    \hline
HfF$^+$ & 48 & 25.8  & 22.8  & 22.5  & 22.5\cite{Skripnikov17}/22.7\cite{Fleig17a} \\
WC      & 48 & 72.8  & 43.5  & 37.9  & 36\cite{Lee13}          \\
TaN     & 48 & 59.3  & 39.6  & 34.8  & 34.9\cite{Skripnikov15a}/36.0\cite{Fleig16} \\
ThF$^+$	& 62 & 42.7	& 36.6	& 36.6  & 37.3\cite{Skripnikov15}/35.2\cite{Denis15} \\
ThO	    & 62 & 98.2	& 83.3	& 79.8  & 75.2\cite{Fleig14}/79.9\cite{Skripnikov16} \\
BaF  	& 30 &  6.5	&  6.4	&  6.3	& 6.52\cite{Talukdar20}/   \\
BaOH	& 30 &  6.5	&  6.5	&  6.4	& 6.4\cite{Denis19}/6.2\cite{Gaul20}     \\
YbF	    & 48 & 22.9	& 23.5	& 23.7	& 23.1\cite{Abe14}      \\
YbOH	& 48 & 22.9	& 23.7	& 24.0  & 23.4\cite{Denis19}/23.8\cite{Prasannaa19}/17.7\cite{Gaul20}   \\
RaF	    & 48 & 55.0	& 54.9	& 54.2	& 52.9\cite{Kudashov14}/52.5\cite{Sasmal16}/50.9\cite{Gaul20a}	          \\
RaOH	& 48 & 54.9	& 55.2	& 54.5  & 52.3\cite{Gaul20}	       \\
HgF	    & 48 &132.0	&118.7  &113.0  & 115.42\cite{Prasannaa15}/116.37\cite{Abe18}     \\
    \hline \hline
  \end{tabular}
  \end{center}
  \vspace{-20pt}
\end{table}

 \begin{table}
  \begin{center}
  \caption{The $|\mathcal{E}_\mathrm{eff}|$ values (GV/cm) from the 
  X2CAMF-HF, CCSD, and CCSD(T) calculations using the analytic energy-gradient theory.
  ``fc ele'' refers to the number of core electrons kept frozen together with virtual orbitals
  higher than 1000 Hartree in
  the CC calculations.}
  \label{tab2}
  \begin{tabular}{ccccccccc}
    \hline \hline
     & fc ele & HF & CCSD & CCSD(T)   \\
    \hline
BaOCH$_3$	& 28 &  6.5	     &        6.4	&6.3	         \\
YbOCH$_3$	& 50 &  22.9     &        23.6	&24.0	         \\
RaOCH$_3$	& 46 &  55.0     &        55.0	&54.2	         \\
LuO	        & 28 &  36.1     &        33.7	&32.4	         \\
LuOH$^+$    & 28 &  32.3     &        29.8	&29.2	         \\
NoF	        & 70 &  185.4    & 	      192.4	&191.9	         \\
NoOH	    & 70 &  185.2    & 	      192.4	&191.7	         \\
LrO	        & 70 &  303.3    & 	      263.9	&246.5	         \\
LrOH$^+$    & 70 &  268.6    & 	      259.5	&255.1           \\
    \hline \hline
  \end{tabular}
  \end{center}
    \vspace{-20pt}
\end{table}

A widely used approach to assess the accuracy 
of the computed $|\mathcal{E}_\mathrm{eff}|$ values is to
compare other properties computed using the same method 
with the corresponding measured values.
We have shown that the X2CAMF-CCSD(T) method provides accurate electric dipole moments and nuclear quadrupole-coupling constants
for heavy-element containing molecules
\cite{Liu18,Liu21}, 
and we expect it to provide accurate $|\mathcal{E}_\mathrm{eff}|$ values. 
Taking advantage of the present extensive benchmark set,
we also take complementary approaches to analyze the accuracy of the X2CAMF-CCSD(T) calculations, 
by comparing the results with
available calculations 
and by analyzing the remaining errors in 
the treatments of relativistic, electron-correlation, and nuclear-model effects. 

The X2CAMF-CCSD(T) $|\mathcal{E}_\mathrm{eff}|$ results agree well with the previous relativistic CC and CI calculations, which are available for all the molecules in Table \ref{tab1} except RaOH,
with the discrepancies amounting to up to several percent of the total values.
The present results for RaOH are consistent with
a recent relativistic density-functional theory calculation. 
It has been reported that the X$^2\Sigma$ ($4f^{14}6s^1$) states of YbF and YbOH 
can be perturbed by the $4f^{13}6s^2$ configuration in CC calculations \cite{Gomes10,Pasteka16}, 
because 
the errors in the treatments of electron-correlation and basis-set effects 
both lead to underestimation of the relative energies 
between the $4f^{13}6s^2$ states
and the X$^2\Sigma$ states \cite{Dolg92}.
It thus is necessary to use large basis sets to obtain
accurate electron-correlation contributions \cite{Prasannaa19,Denis19}.
We note that the present Kramers unrestricted calculations yield
smaller electron-correlation contributions for YbF and YbOH than in previous studies \cite{Abe14,Prasannaa19},
while the total CCSD values agree well with those in Refs. \cite{Abe14,Prasannaa19,Denis19}. 
The present CCSD result for YbOH also agree well with
the Fock-space CCSD value in Ref. \cite{Denis19}. 
It has been shown that the CCSD(T) noniterative triples corrections to first-order properties of Yb-containing are not accurate representation of triples corrections \cite{Pasteka16}. The CCSD(T) triples corrections for YbF, YbOH, and YbOCH$_3$ in Tables \ref{tab1} and \ref{tab2} thus only serve as rough estimates for triples contributions in the present discussion. 

Table \ref{tab2} summarizes calculations for nine molecules,
for which no calculations have been reported. 
The metal methoxides, BaOCH$_3$, YbOCH$_3$, and RaOCH$_3$ 
possess $|\mathcal{E}_\mathrm{eff}|$ values similar 
to those of the corresponding fluorides and hydroxides. 
For example, the X2CAMF-CCSD(T) $|\mathcal{E}_\mathrm{eff}|$ value for YbOCH$_3$ amounts to 24.0 GV/cm, 
very similar to the values of 23.7 and 24.0 GV/cm for YbF and YbOH. 
This is consistent with the chemical intuition that the unpaired electron in YbOCH$_3$ is localized at the Yb atom. 
These computational results support the recent proposal of using the nearly degenerate rotational states of 
  these symmetric-top molecules to improve the sensitivity of eEDM measurements \cite{Kozyryev17}.
The $|\mathcal{E}_{\text{eff}}|$ values of 32.4 GV/cm and 29.0 GV/cm for LuO and LuOH$^+$ 
are a little larger than those of YbF and YbOH and are similar to that of ThF$^+$. 
The four small molecules containing late actinide elements, 
NoF, NoOH, LrO, and LrOH$^+$, exhibit particularly large $|\mathcal{E}_{\text{eff}}|$ values of 191.9, 191.7, 246.5, and 255.1 GV/cm, respectively, because of the relativistic enhancement in the presence of these very heavy atoms. 
 
  \begin{table}
  \begin{center}
  \caption{The errors of the X2CAMF scheme, the core-correlation contributions, and the finite nuclear size effects
  for $|\mathcal{E}_\mathrm{eff}|$ (GV/cm). The percentages of the total values are enclosed in the parentheses. }
  \label{tab3}
  \begin{tabular}{ccccccccc}
    \hline \hline
     &  \multicolumn{1}{c}{X2CAMF error$^a$}  & \multicolumn{2}{c}{Core correlation$^b$} & \multicolumn{1}{c}{Finite nuclear size$^c$} \\
    \hline
    & HF & fc ele & CCSD & CCSD  \\
    \hline
BaF	    &  -0.03 (-0.5\%)    & 0	& 0.01 (0.2\%)	 &-0.04   (-0.5\%) \\
YbF	    &  -0.01 (-0.0\%)    &28	&-0.06 (-0.2\%)	 &-0.38   (-1.6\%) \\
HfF$^+$ &   0.01 (0.0\%)     &10	& 0.40 (1.7\%)	&-0.47   (-2.0\%) \\
RaF	    &   0.14 (0.3\%)     &28	& 0.12 (0.2\%)	 &-3.67   (-6.6\%) \\
ThO	    &   0.31 (0.3\%)     &28	& 0.25 (0.3\%)	 &-6.27   (-7.4\%) \\
ThF$^+$	&   0.50 (1.2\%)     &28	& 0.48 (1.3\%)   &-3.10   (-8.2\%) \\
NoF	    &   0.95 (0.5\%)     &48	&-0.30 (-0.2\%)	 &-35.04 (-18.0\%) \\
    \hline \hline
  \end{tabular}
  \end{center}
  a. The differences between the Dirac-Coulomb-Gaunt and X2CAMF results. 
       b. The differences between the present calculations (freezing the "fc ele" number of core electrons and
  virtual orbitals higher than 2000 Hartree) and those in Tables \ref{tab1} and \ref{tab2}.
  c. The differences between the results using the Gaussian and point-like nuclear models. 
    \vspace{-10pt}
\end{table}

The computational efficiency of the X2CAMF scheme
originates from the use of the untransformed two-electron Coulomb interaction
and the AMF approximation to the two-electron spin-orbit integrals and the Gaunt term.
As shown in Table \ref{tab3},
the errors are small across the periodic table, e.g, the error amounts to -0.5\% for BaF, 0.3\% for RaF, 1.2\% for ThF$^+$, and 0.5\% for NoF.
We mention that the Gaunt-term contributions
amount to around 1\% for all the molecules studied here, except that it is around 3\% for HfF$^+$. We expect the remaining relativistic contributions 
from the Gauge term and quantum electrodynamics
to be smaller than the Gaunt-term contributions. 


The differences between the CC and HF results in Tables \ref{tab1} and \ref{tab2} 
represent the electron-correlation contributions. 
The electron-correlation contributions amount to
more than half of the total values for TaN and WC. 
In contrast, they are less than 2\% for MF, MOH, MOCH$_3$, with M=Ba, Yb, Ra. 
The rest of the molecules receive moderate yet important contributions from electron correlation, 
ranging from 5 to 30\%.
The magnitude of triples contributions
are significantly smaller than singles and doubles contributions for most of the molecules studied here,
except for some molecules exhibiting very small total correlation contributions such as BaOH, RaF, RaOH, and RaOCH$_3$. 
WC and TaN exhibit relatively large triples contributions of around 10\%. 
We expect the high-level correlation contributions 
to be smaller than triples contributions. 
Finally, as shown in Table \ref{tab3}, the correlation of the inner-shell core electrons 
makes minor contributions, amounting to up to a few percent.


The importance of
the nuclear model increases rapidly for heavier elements, as demonstrated in Table \ref{tab3}. 
While the difference between the Gaussian model 
and point-like model is only -0.5\% for BaF, it amounts to -7\% 
for ThO and -18\% for NoF. 
Since the Gaussian model is more realistic than the point-like model, 
we expect the errors of the Gaussian model 
to be much smaller than the difference between the Gaussian and point-like models. 

Taking these error analyses into account, we conclude that the computational results in Tables \ref{tab1} and \ref{tab2} are accurate to within 10\% except that we assign a 25\% error estimate for NoF, NoOH, LrO, and LrOH$^+$. Further improvement of the error estimate requires a study of the sensitivity of computed results to the sizes and function form of the finite nuclear model. 
The current conservative error estimate still supports that the molecules containing late actinide elements possess extraordinarily large $|\mathcal{E}_{\text{eff}}|$ values. Therefore, given the structure of a molecule, the present analytic-gradient based scheme only needs a single X2CAMF-CCSD(T) analytic-gradient calculation correlating valence and semicore electrons, which is of black-box nature, to provide a $\mathcal{E}_\mathrm{eff}$ value accurate enough for the initial screening of candidate molecules for eEDM measurements. One may improve the results by correcting the errors of the X2CAMF scheme and by including inner-shell correlation (Table \ref{tab3}), e.g., the best values for YbF, HfF$^+$, ThO, and ThF$^+$ from the present calculations are obtained as 23.4, 22.9, 80.4, and 37.6 GV/cm, respectively, by combining the small corrections in Table \ref{tab3} with the CCSD(T) results in Table \ref{tab1} for all these molecules except that we adopt the CCSD result for YbF.

\section{Summary and Outlook}
 
  We report the development of an analytic-gradient based method for the X2C-CCSD(T) calculations of effective electric field in paramagnetic molecules. Extensive benchmark calculations demonstrate the efficiency and accuracy of the present scheme. The extension of the present method to the calculations of other symmetry-violating parameters, e.g., the parameter associated to the measurements of nuclear magnetic quadrupole moment \cite{sushkov1984possibility,kozlov1987calculation,skripnikov2015tan,fleig2016tan,maison2020search,denis2020enhanced}, is straightforward, by contracting the reduced density matrix with the corresponding property integrals. The present method thus provides significantly enhanced capabilities for the calculations of symmetry-violation sensitivity parameters in molecules. It will enable convenient, efficient, and reliable calculations of these parameters to help engineer new molecules 
  suitable for the search of new physics via precision measurement. 
  
  {\textit{Acknowledgement ---}} 
  L. C. is grateful to Nicholas Hutzler (Pasadena), Phelan Yu (Pasadena), Kia Ng (Boulder), and Anastasia Borschevsky (Groningen) for helpful comments on the manuscript and for stimulating discussions. 
  This work is supported by 
  the National Science Foundation, under grant No. PHY-2011794.
  All calculations were carried out at the Maryland Advanced Research Computing Center (MARCC).
 
  C. Z and X. Z. contributed equally to the work. 
  
\bibliography{uo2cl2}


\end{document}